# Foo Fighters and Microwave Oven Plasma Balls

Chris Allen Broka
(chris.broka@gmail.com)

**Abstract.**

An explanation is given for two peculiar phenomena ─ Foo Fighters and microwave oven plasma balls. Both are proposed to arise from the interaction of microwave energy with areas of hot, partially ionized, gas. In the former case it is suggested that the airplane's own radar irradiated its hot exhaust thus producing a glowing object that appeared to follow it. The physics of these situations is examined and we attempt to delineate the temperatures and field strengths necessary for such processes to be viable. To do this we employ an interesting iterative method for the solution of the appropriate partial differential equations. We find that both processes do make good physical sense although the field strengths required are towards the higher end of what would be considered reasonable.

## 1. Introduction.

The mysterious appearance of "Foo Fighters" over the skies of World War II, and their equally mysterious disappearance after the cessation of hostilities, constitutes one of the stranger episodes in the history of atmospheric physics (1). Bright lights were reported to follow aircraft. They were observed in all theaters of the war. Common opinion on most sides held that they were enemy secret weapons or UFOs. Little evidence indicating this surfaced following the end of the war and they were variously explained as corona discharges, ball lightning, or other electromagnetic or optical phenomena.

The idea that these lights were produced by radar beams interacting with ionized gas, particularly engine exhaust, is not new. An interesting video suggesting as much is easily found on the Internet (2). But the provenance of this notion is difficult to trace and it does not seem to have attracted much physical attention. For a phenomenon such as that described to have been real it would have had to have an input of energy (or be an actual UFO). Very few sources of external energy, save radar energy emitted by the aircraft itself would have been present to account for a glowing object. We would like to investigate this idea and find out whether, and under what circumstances, it might be plausible.

## 2. Microwave Oven Plasma Balls.

Most people that would read a paper like this have, probably, at some time or other, succumbed to the temptation to put a lit candle or sliced grape into their microwave oven. The result is usually a bright plasma ball that rises to the top surface of the oven and dissipates. What could be occurring here? Surprisingly, little theoretical work seems to have gone into analyzing this common phenomenon. Das (3) has tried to describe it in terms of Townsened avalanches (although the electric field strengths involved do not seem to easily justify this). We will take a far simpler approach.

Suppose there is a free electron oscillating in a microwave field. It moves according to:

1) $m_e\, x''(t) + \lambda\, x'(t) = e\, E\, \mathrm{Sin}(\omega\, t) = F(t)$

where $\lambda$ represents a damping constant that describes the interaction of the electron with the air it is moving through. We can estimate $\lambda$ by considering Paschen's work on the electrical breakdown of air at STP (4). He finds that the average distance traveled between electron collisions with air molecules is, roughly, $10^{-6}$ m. Knowing this we can estimate $\lambda \approx 10^{-19}\ \frac{\text{kg}}{\text{sec}}$. Realizing that the electron gives energy to its surrounding gas at a rate x'(t) F(t) we find that the gas takes up energy at a rate:

2) Rate energy in (W/$m^3$) = $\frac{\eta_e\, e^2\, E^2\, \lambda}{2\left(m_e^2\, \omega^2 + \lambda^2\right)} \approx \frac{\eta_e\, e^2\, E^2}{2\, \lambda}$ where $\eta_e$ is the number density of free electrons. $\omega$ is assumed to be about $10^9$-$10^{10}$ $\text{sec}^{-1}$ since this is typical for a kitchen microwave oven.

Now, assuming that each gas molecule possesses an energy $\beta$ k T, we can write the rate of temperature increase as:

3) $\partial_t\, T(t) = \frac{\eta_e\, e^2\, E^2}{2\, \eta_A\, \beta\, k\, \lambda}$ where $\eta_A$ represents the number density of air molecules. $\beta$ is set at 3.5, as is proper for diatomic molecules at high temperature.

The pressure of the air is taken to always be 1 atm ($P_0$ = 101,325 N/$m^2$). We will picture our nascent plasma ball having a radius $d_0$ and containing a fixed number of free electrons. We cannot be sure what this number is but, for simplicity's sake, we will assume their total number to be more-or-less constant. We will assume that all the air molecules inside our plasma ball stay there forever and we will write their number as $N_0$. As our ball heats it will, of course, expand. $\eta_e$ and $\eta_A$ will change equally. $\lambda$ may be designated $\lambda_0 T_0$/T(t) where $T_0$ is about 273 K. (As the ball expands the electrons will collide less often and $\lambda$ will decrease as 1/volume.) Therefore:

4) $\partial_t\, T(t) = \frac{\eta_e\, e^2\, E^2\, T(t)}{2\, \eta_A\, \beta\, k\, \lambda_0\, T_0}$.

It is not clear what $\eta_e$ really is. But we know that there are free electrons in candle flames since these can conduct electricity somewhat. We do know that a candle flame can easily be up to 1800 K in places, however. $E$ is more difficult to estimate because of the complexity of the situation within the oven. A pattern of standing waves is set up and the field can be very strong in some places and weak in others. We know that a microwave oven projects about 1 kW of energy through a roughly .07 $m^2$ area. So $E \approx$ 3250 V/m on average. The intensity of this radiation would be about 14,000 W/$m^2$. One of two things can happen. If $\eta_e$ is very small, nothing happens. If $\eta_e$ is sufficiently great the little ball will heat exponentially in real time. If $\eta_e$ = 6.5 X$10^{16}$ $m^{-3}$ (corresponding to about 1.5X$10^{-6}$ % ionization) it will heat to 8, $000^o$ in 1 sec! But we should also bear in mind that, as the temperature increases, $\eta_e$ will increase due to simple (Saha) ionization of the air. The net result would be a rapid expansion of the ball to enormous temperatures and dimensions. Obviously, no such catastrophes ever happen. Why not? For one thing, the plasma ball (being very hot and buoyant) quickly rises to the top of the oven compartment where it dissipates. But, more importantly, there is only so much energy an oven can give to such a ball. Suppose our ball of candle flame began with a radius of 2 cm. If $\eta_e$ exceeds a critical level the above model would have it taking up energy at a rate greater than $I_0$ Area = $I_0\ \pi\ (2\ \text{cm})^2$ where $I_0$ designate the radiation intensity impinging on the ball. This is, of course, quite impossible. Since $I_0$ = $\frac{\epsilon_0\, c\, E^2}{2}$ we see that the critical electron density is defined by:

5) $\eta_e \geq \frac{3 \lambda_0 c \epsilon_0 T_0}{4 e^2 T d}$ which is independent of $I_0$. $T_0$ = 300 and $\lambda_0$ is the damping at STP.

This gives a critical density of 6.5X$10^{16}$ $m^{-3}$ for a 2 cm ball and 1.3X$10^{16} m^{-3}$ for a larger one with a 10 cm radius, assuming these are at 1800 K. It is perhaps worth mentioning that, at these and higher electron densities, the plasma frequency is generally above the frequencies used by microwave ovens (and radars). Early in its life we suppose these free electrons are supplied primarily by the chemical reactions that power the flame. With the temperature above about 4000-5000 K Saha ionization of the air would provide more than enough electron density to sustain the process.

If $\eta_e$ is above the critical value the above-described heating mechanism cannot work and must be abandoned in favor of one where the heating occurs at a slower rate given by:

6) $\partial_t T(t) = \frac{\pi d^2 I_0}{N_0 \beta k}$ where $d$ is the radius of the expanding ball of ionized air and $N_0$ is the number of air molecules inside it.

This radius will be proportional to $(T(t) / T_B)^{1/3}$ where $T_B$ is the initial temperature of the flame. Thus $\partial_t T(t) \propto T(t)^{2/3}$ and the ball will expand less quickly than it would according to Eqn 4). Indeed, we will suppose that $\eta_e$ always exceed the critical value. This assumption greatly simplifies our work. We write:

7) $\partial_t T(t) = \frac{\pi d^2 I_0}{N_0 \beta k}$ inside the ball and 0 outside.

But the temperature will also change owing to thermal diffusion. And, if hot enough, the ball (which we assume to be optically transparent) will cool owing to its emission of continuum radiation. Therefore:

8) $\partial_t T(t, r) = \frac{\pi d^2 I_0}{N_0 \beta k} + D_0(T(t, r)) \nabla^2 T(t) - \frac{W(T(t, r))}{P_0 \beta}$ where $W(T(t, r))$ represents the loss to radiation ($\frac{W}{m^3}$) and $D_0(T(t, r))$ the thermal diffusivity of the air ($\frac{m^2}{\sec}$). The source term is understood to operate only inside the ball.

## 3. Numerical Simulations.

In order to see whether Equation 8) describes any reasonable physics it must be solved numerically. The solution of this non-linear PDE is not straightforward. It is difficult, using *Mathematica*'s NDSolve function, to capture the expansion of the ball. This is a crucially important matter since it directly influences the rate of the ball's heating. We require a more effective method.

We will construct a table, *K*, that represents the initial state of the ball:

```
K = {{0.001, 1800}, {0.002, 1800}, {0.003, 1800}, {0.004, 1800}, {0.005, 1800}, {0.006, 1800},
    {0.007, 1800}, {0.008, 1800}, {0.009, 1800}, {0.01, 1800}, {0.011, 1800}, {0.012, 1800},
    {0.013, 1800}, {0.014, 1800}, {0.015, 1800}, {0.016, 1800}, {0.017, 1800}, {0.018, 1800},
    {0.019, 1800}, {0.02, 1800}, {0.021, 300}, {0.022, 300}, {0.023, 300}, {0.024, 300},
    {0.025, 300}, {0.026, 300}, {0.027, 300}, {0.028, 300}, {0.029, 300}, {0.03, 300}, {0, 0}}.
```

Here the first term in the elements of the set designate the radial distance from the center of the ball and the second its temperature. We imagine that the ball begins at 1800 K – not too unbelievable – and that the air outside is at 300 K. Essentially, we are treating the ball as a sort of "onion." We assume that nothing ever crosses from one onion-layer to another. Any air molecule or free electron that starts in a given shell stays there forever. Heat is only given to the innermost 20 shells. The 31st element of the set is a 'dummy element' that is only used to index time throughout the simulation. We guess that the initial radius is 2 cm. This may seem a little large for a candle flame. But we should recall that a significant amount of very hot air surrounds what our eye sees as the flame. Smaller flames can be modeled. They do not, however, achieve a size or brightness consistent with what we usually see in our ovens.

We first act on *K* with a 'Source Operator' which adds the proper amount of temperature to each layer of the onion and subtracts the loss due to continuum radiation (usually insignificant). After this a 'Diffusion Operator' acts upon it doing two things. Firstly, it redistributes the temperatures according to ordinary thermal diffusion. Then it adds one time increment to the 'dummy' 31st set. Lastly an operator rearranges the coordinates of the onion-layers so that expansion is accounted for – if, say, the temperature inside a particular layer happened to increase, the separation between that layer's walls would have to increase as well. These operators execute over a time step small enough that the solution is stable and accurate. We combine these three sequential processes into one 'Evolution Operator' which is iterated upon K as many times as necessary to cover the time interval of interest. The details are laid out in the Supplementary Material. Having estimated the candle flame as being about 1800 K we have only to figure out $I_0$. The pattern of microwaves inside a typical oven has been simulated (5). The strongest fields can exceed 25,000 V/m. Also, as standing waves are set up, energy can impinge on the ball from both sides. We will guess that $I_0 = 10^6$ W/$m^2$. This may be towards the high end of what might be considered reasonable. But it gives us an encouraging result (Figs. 1, 2).

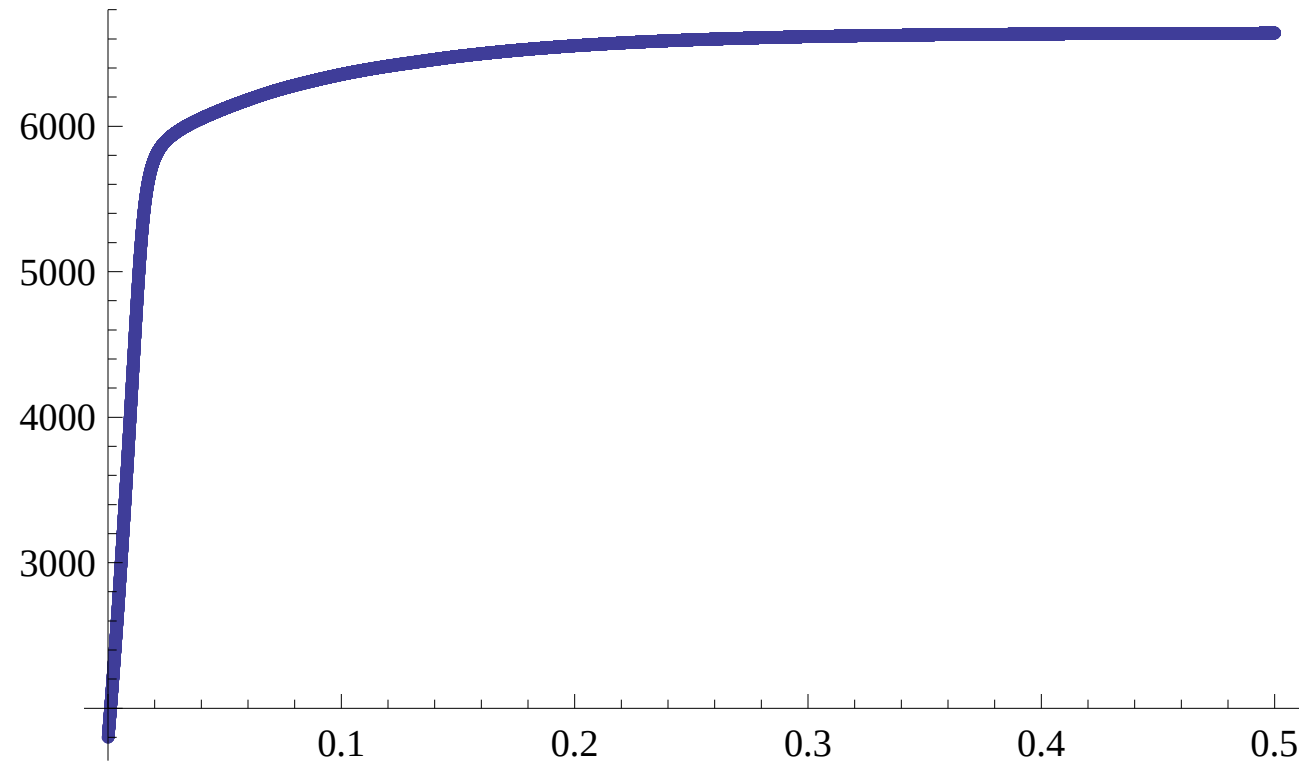

Fig. 1

The calculated core temperature (*K*) of an $d_0 = .02$ m, $I_0 = 10^6$ W/$m^2$ ball as a function of time (sec).

The ball quickly heats and expands which is exactly what we see in our ovens at home. By the time it reaches about 3500 K it will be quite visible. By the time it reaches 6000 K it will be glowing with the apparent intensity of an 45 Watt light bulb (assuming these are about 2% efficient (6)). Below we plot its temperature profile at t = .5 sec (Fig. 2). The ball achieves a maximum radius of about .03 *m*.

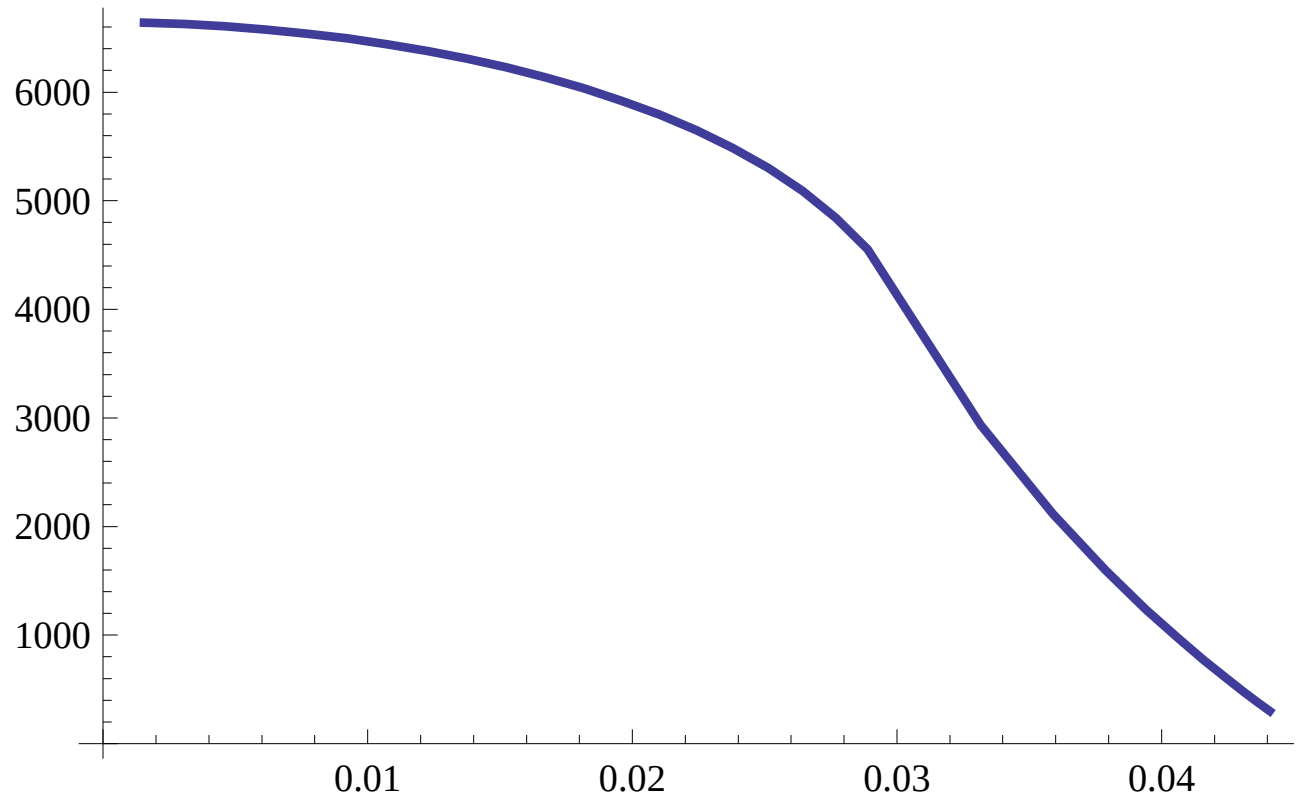

Fig. 2

The temperature profile (*K*) of an $d_0$ = . 02 m, $I_0 = 10^6$ W/$m^2$ ball at t = .5 sec as a function of radial distance (*m*).

## 4. Foo Fighters.

Let us apply what we have learned above to the, more interesting, problem of Foo Fighters. Situations in the skies over World War II differed quite a bit from those inside a microwave oven. But we do know that airplanes put out hot exhaust. Modern propeller-driven planes can put out exhaust at, at least, about 1500 K (7). We will guess that the powerful, early, war planes put out exhaust at 1800 K (since we have, conveniently, already used this number and, besides, it is as reasonable as any).

The primitive airborne interception radars in use at the time could generate as much as 250 kW in power (8). Some focused it using parabolic reflectors. Others utilized Yagi-Uda or other dipole antenna arrays (9). Not knowing the specifics, we will estimate that 250 kW went out over an area of roughly 1 $m^2$. (It cannot, really, have been much larger or smaller for things to make practical sense.) These early radars did not pour their energy out in a continuous stream but, rather, delivered it in pulses. The pulse rate frequency could be as high as many hundred Hz. But the German Hohentwiel system operated at as low a frequency as 50 Hz (10). We will just use 100 Hz as an illustrative and reasonable example. Different planes had, of course, different radar sets on board which they would use to scan for enemy aircraft. If a radar beam were to end up focused upon an area where the airplane was discharging hot, probably rather ionized, exhaust, a phenomenon similar to that already described might very well take place. A hot plasma ball might form which, for as long as it remained subject to the plane's radar, would grow and luminesce. Since an airplane's exhaust stream is much larger than a candle flame we will consider a ball with $d_0$ = .1 m. Below we plot the results (Fig. 3).

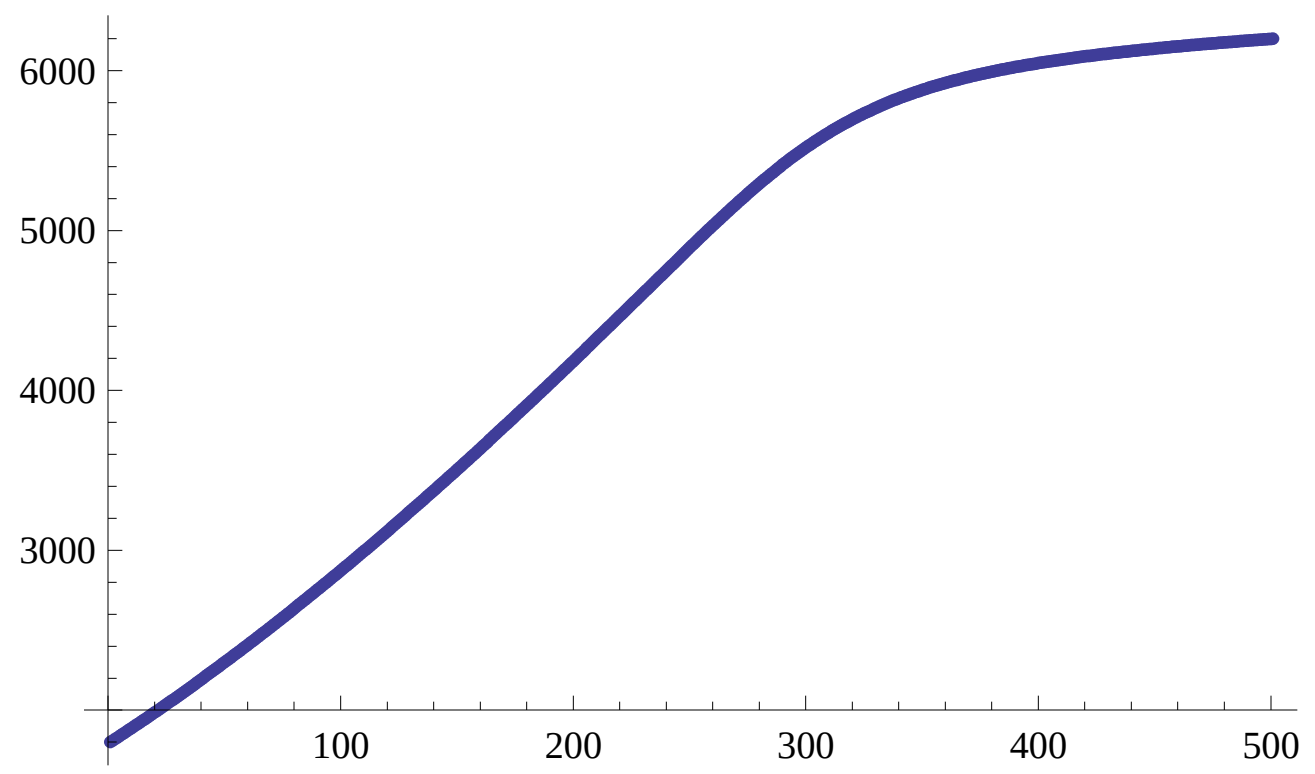

Fig. 3

The calculated core temperature $(K)$ of an $d_0$ = . 1 m, $I_0$ = 2.5 X$10^5$ W/$m^2$ ball as a function of time (msec).

A concern may arise since the radar supplies energy to the ball in pulses whereas we are approximating it as a continuous stream of power. Simulations have been performed with the source turning off and on rapidly. Although the temperature increases in a more jagged way the essential conclusions are unaltered (data not shown). It is remarkable how both this system, and its far smaller, oven-based, counterpart converge to the same final temperature of about 6300 K. Once fully developed the plasma ball would glow with the apparent brightness of a roughly 660 W incandescent light bulb. Now this is a bit below what we might like to see for a Foo Fighter. So we do not think this is all there is to it.

Every .01 sec the exhaust stream will receive a powerful pulse of microwave energy. World War II airplanes generally flew at about 100 m/sec. The result would be a string of plasma balls that trailed behind the plane. They would be spaced at 1 m intervals. As long as they remained in the radar beam they would continue to heat and grow. The ones closest to the plane would be small and cool. Those farther away would be large and bright. Obviously, such balls could not continue to grow forever; at some point one of three things would happen: 1) The ball would simply convect away and dissipate. 2) The radar beam might move away from it thus depriving it of sustenance. Or, 3) so many balls might trail the plane that they, together, would absorb all the energy put out by the transmitter. We cannot say much about the first two possibilities. But we know that one of these balls has an area of about .03 $m^2$. If all the emitted energy from the plane covers a rough effective area of 1 $m^2$ the maximum number of balls would be about 1/.03 = 33. We will use 35 since it is a more convenient number. So, out to about 35 m behind the plane, the air crew would see a string of lights with the brightest being the farthest away. Beyond 35 m the lights would quickly die out and vanish. We plot this situation below (Fig. 4).

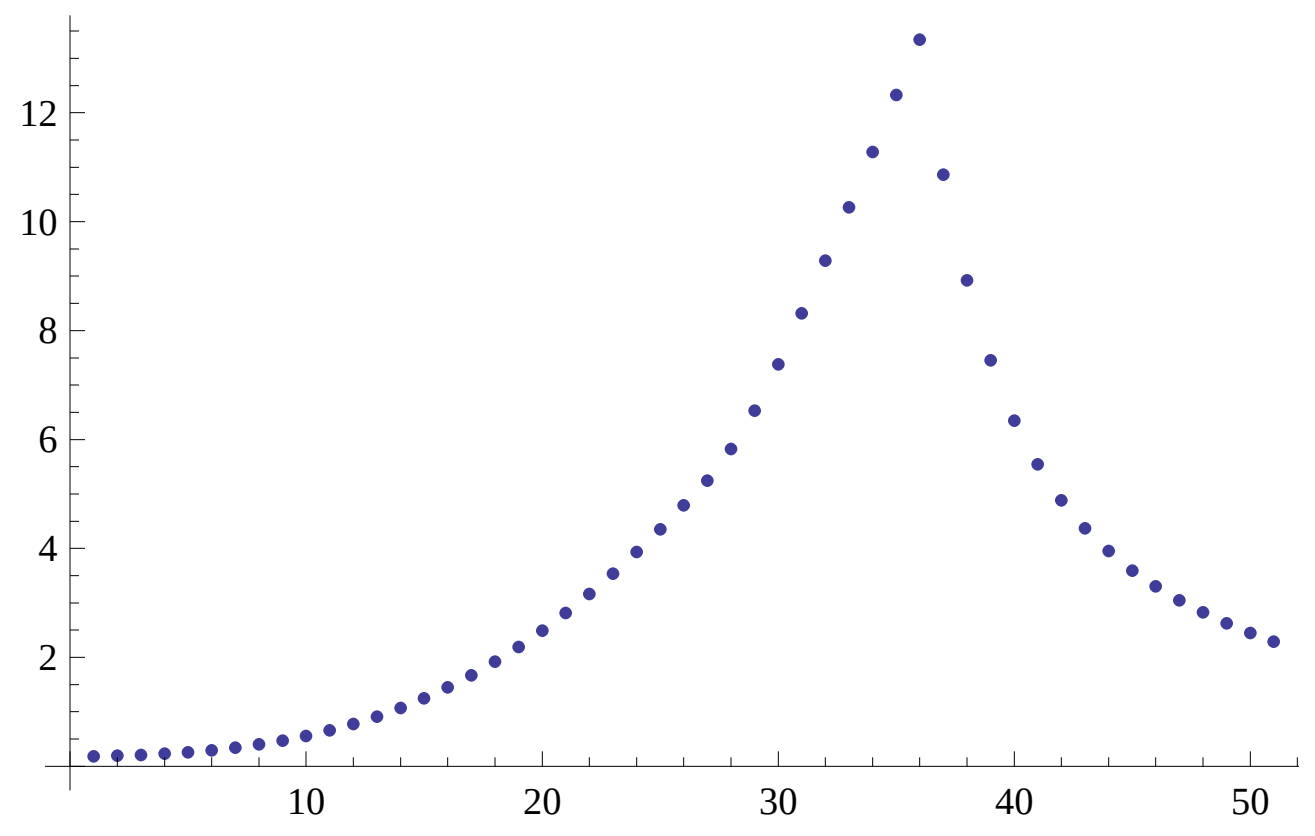

Fig. 4

The luminosity (W) of a string of $d_0 = .1$ m, $I_0 = 2.5\ \mathrm{X}10^5\ \mathrm{W}/m^2$ balls as a function of distance from aircraft (m).

So we do not think the air crews were observing a *single object*. Instead, they were looking at something like the still frames in a movie. The balls closest to the plane would be easily overlooked. They would probably focus their attention on the ~5 very bright balls that would form a chain about 4-5 m long and 35 m behind the plane. These would glow, collectively, with about 55 W (or the intensity of a 2750 W light bulb). Distance and confusion could easily lead them to perceive a single, large and bright, object that appeared to be chasing their plane. We suspect that this is what happened. The picture is, probably, complicated somewhat by the fact that not all balls in such a string would have ignited or behaved identically.

It could be objected that we have had to use some pretty large values for $I_0$ and $T$ both for the Foo Fighters and the microwave plasma balls. Our values, while not physically ridiculous, are, to be sure, "generous." But Foo Fighters were not observed chasing all airplanes every day and not all candles perform well in the oven. Some rather fortuitous circumstances seem required for these phenomena to work. How a powerful radar beam would be "painting" the balls behind an aircraft is hard to say. The previously mentioned video blames 'overpowered' radar sets. This is certainly possible – malfunctions must have been common. Many airplanes had radar that could scan off to the sides and behind so it may be that some air crews exacerbated their own problem by trying to track the, seemingly dangerous, object that was pursuing them. In the process they would only have encouraged its growth and bad behavior! We also wonder why Foo Fighters largely disappeared following the end of the War. This may be due to the adoption of jet power – perhaps jet engines do not produce suitable exhaust. Also, radar systems evolved becoming much more efficient and manageable.

## 5. Conclusion.

We have tried to show that the interaction of microwave energy with hot, ionized, air can very well produce the kinds of phenomena encountered in our kitchen ovens and by World War II fliers. We have been able to estimate the circumstances that seem necessary and find these to be, generally, not too physically unrealistic. It may be objected that our treatment of the heating mechanism for the ball is a bit simplistic and threadbare. This is true. But the important point to remember is that the ball can never take up more energy than is incident upon it. We just assume it does the very best it can.

## References.

# Supplementary Material - Microwave Oven Plasma Balls.

**Below we define W[T]. Our values are taken from (Lowke, J.J., M.A.Uman, R.W.Liebermann, 1969 : Toward a theory of ball lightning. *J. Geophys. Res*., 74, 6887 - 6898.) and (Yos, J.M., 1963 : Transport properties of nitrogen, hydrogen, oxygen, and air up to 30 000 K. *AVCO Tech. Memo*. RAD - TM - 65 - 7, AD - 486 068, doi : http : // www.dtic.mil / dtic / tr / fulltext / u2 / 435 053. pdf.).**

```
w1 = Interpolation[{{0, 0}, {1000, 1.5}, {1500, 4.5}, {2000, 15}, {3000, 150},
   {5000, 1500}, {10 000, 7.47 × 10^7}, {12 500, 10^9}, {20 000, 7 × 10^9}, {30 000, 10^10}}]
```

```
InterpolatingFunction[{{0., 30 000.}}, <>]
```

```
w2[t_] := Exp[.00216329 t - 3.50388]
```

```
W[t_] := If[5000 < t < 10 000, w2[t], w1[t]]
```

**Here we define the ball' s size and chacteristics :**

```
d0 = .02
k = 1.38 × 10^-23
β = 3.5
η0 = 2.45 × 10^25
T0 = 1800
P0 = 101 325
I0 = 1 000 000
```

**Here is the number of air molecules inside our ball.**

```
N0 = 4 (300 / 1800) Pi η0 d0^3 / 3
```

$1.36834 \times 10^{20}$

**Here is its volume.**

```
V0 = 4 Pi d0^3 / 3
```

0.0000335103

**This is the number density of air molecules inside the ball.**

```
N0 / V0
```

$4.08333 \times 10^{24}$

**This is the heat absorbed over a unit area.**

```
Pi I0 / (N0 k 3.5)
```

$4.75345 \times 10^{8}$

**HH simply diffues a solution according to $\nabla^2 T(t)$.**

```
HH[U_] :=
 Table[{i, Which[i == 1, 2 (U[[2]][[2]] - U[[1]][[2]]) / U[[1]][[1]]^2, i == 30, 0, i == 31, 0, True,
    2 ((U[[i + 1]][[2]] - U[[i]][[2]]) / (U[[i + 1]][[1]] - U[[i]][[1]]) - (U[[i]][[2]] -
           U[[i - 1]][[2]]) / (U[[i]][[1]] - U[[i - 1]][[1]])) / (U[[i + 1]][[1]] - U[[i - 1]][[1]]) +
     2 (U[[i]][[2]] - U[[i - 1]][[2]]) / ((U[[i]][[1]] - U[[i - 1]][[1]]) U[[i]][[1]])]}, {i, 1, 30}]
```

```
G[U_, M0_, i_] := (U[[i]][[2]] / M0[[i]][[2]])
```

**Below we define the thermal diffusivity. These values are derived from (Yos, J.M., 1963 : Transport properties of nitrogen, hydrogen, oxygen, and air up to 30 000 K. *AVCO Tech. Memo*. RAD - TM - 65 - 7, AD - 486 068, doi : http : // www.dtic.mil / dtic / tr / fulltext / u2 / 435 053. pdf.).**

```
D0 = Interpolation[{{300, .000022}, {1000, .00015}, {2000, .00052},
   {3000, .0024}, {4000, .0043}, {5000, .009}, {6000, .027}, {7000, .053},
   {10 000, .024}, {15 000, .065}, {20 000, .093}, {25 000, .195}, {30 000, .3}}]
```

```
InterpolatingFunction[{{300., 30 000.}}, <>]
```

**We use a time step of .00001 sec.**

```
Diff[U_] := Table[If[i ≤ 30, {U[[i]][[1]], U[[i]][[2]] + .00001 D0[U[[i]][[2]]] HH[U][[i]][[2]]},
   U[[i]] + .00001 {1, 1}], {i, 1, 31}]
```

**Below we define the initial state of our candle flame.**

```
K = Table[Which[i ≤ 20, {.001 i, 1800}, 20 < i ≤ 30, {.001 i, 300}, True, {0, 0}], {i, 1, 31}]
```

```
{{0.001, 1800}, {0.002, 1800}, {0.003, 1800}, {0.004, 1800}, {0.005, 1800}, {0.006, 1800},
 {0.007, 1800}, {0.008, 1800}, {0.009, 1800}, {0.01, 1800}, {0.011, 1800}, {0.012, 1800},
 {0.013, 1800}, {0.014, 1800}, {0.015, 1800}, {0.016, 1800}, {0.017, 1800}, {0.018, 1800},
 {0.019, 1800}, {0.02, 1800}, {0.021, 300}, {0.022, 300}, {0.023, 300}, {0.024, 300},
 {0.025, 300}, {0.026, 300}, {0.027, 300}, {0.028, 300}, {0.029, 300}, {0.03, 300}, {0, 0}}
```

**Source describes the heat given to each onion layer.**

```
Source[U_] :=
 Table[Which[i ≤ 20, .00001 {0, If[U[[31]][[2]] ≤ 1, 1, 0] 4.75×10^8} (U[[20]][[1]])^2 + U[[i]] -
    {0, 1} .00001 U[[i]][[2]] W[U[[i]][[2]]] / (P0 β), True, {0, 0} + U[[i]]], {i, 1, 31}]
```

```
HD[U_] := Diff[Source[U]]
```

**FF expands everything according to the changes in temerature. It also redefines the radial coordinates with which we work.**

```
FF[U_, M0_] := Table[{(Σ_{k=1}^{j} (G[U, M0, k] (U[[k]][[1]]^3 - If[k > 1, 1, 0] U[[k - 1]][[1]]^3)))^(1/3),
    U[[j]][[2]]}, {j, 1, 31}] // N
```

```
Evolve[U_] := FF[HD[U], U]
```

**Here we calculate our solution. It can take up to 40 min.**

```
NestList[Evolve, K, 50 001];
```

**CT gives the ball's core temperature.**

```
CT[U_] := U[[1]][[2]]
```

```
Table[CT[%52[[i]]], {i, 1, 50 001}];
```

```
ListPlot[%, PlotRange → All]
```

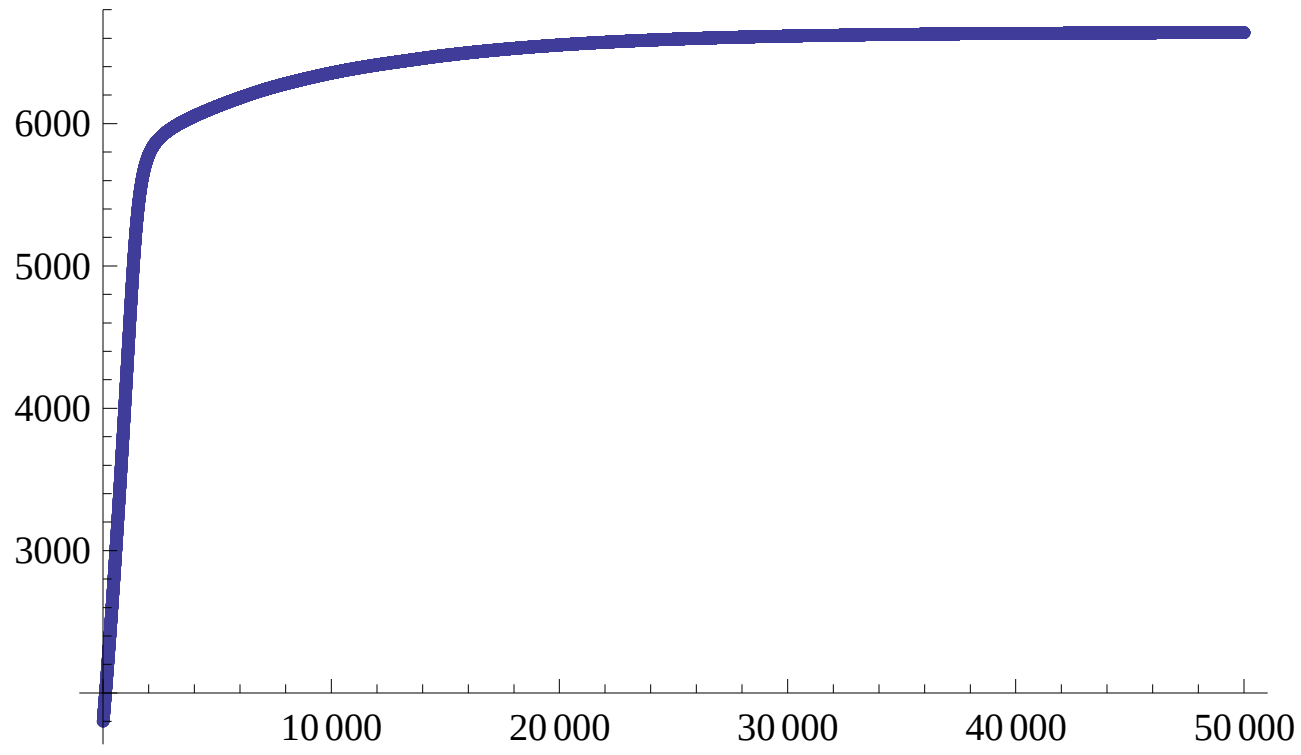

**We can find the values at t = .5 sec and graph them.**

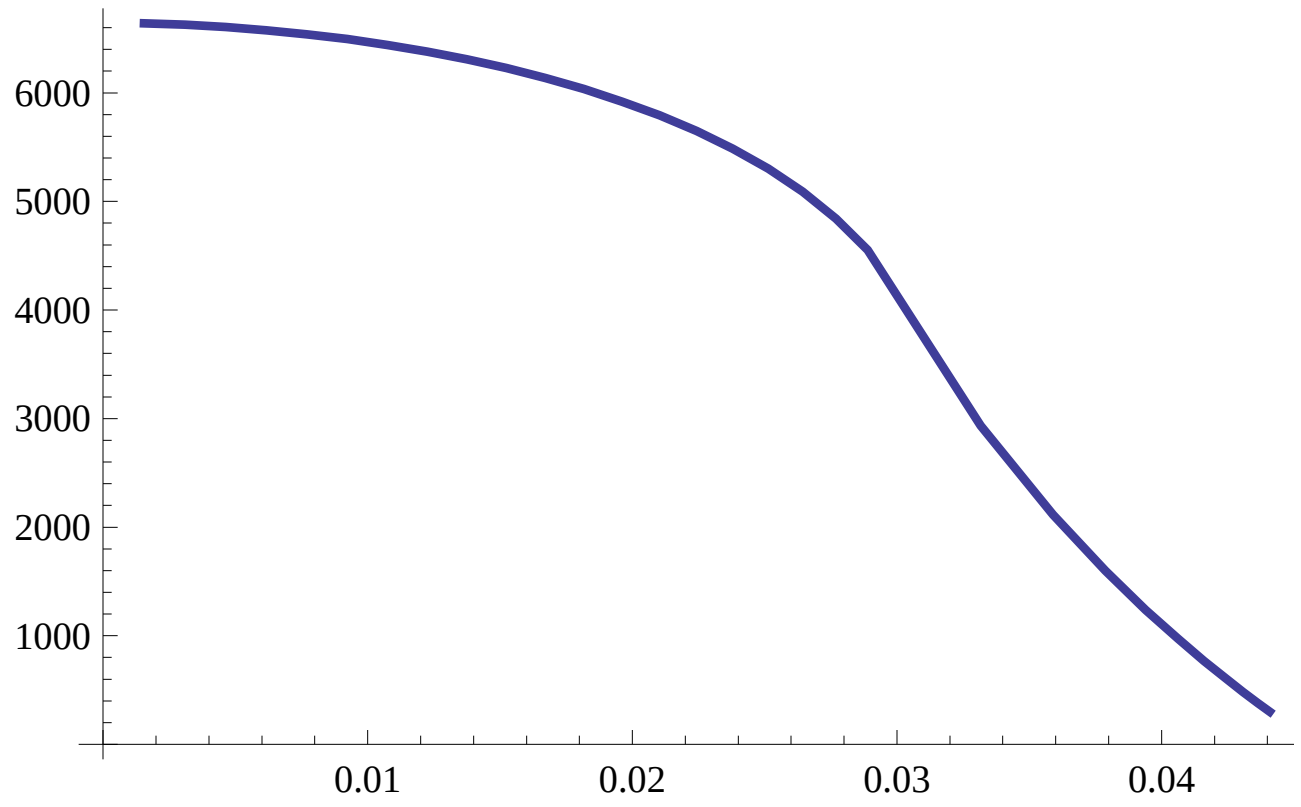

**We can graph the ball' s luminosity.**

```
Table[W[CT[%52[[i]]]], {i, 1, 50 001}];
```

```
ListPlot[%, PlotRange → All]
```

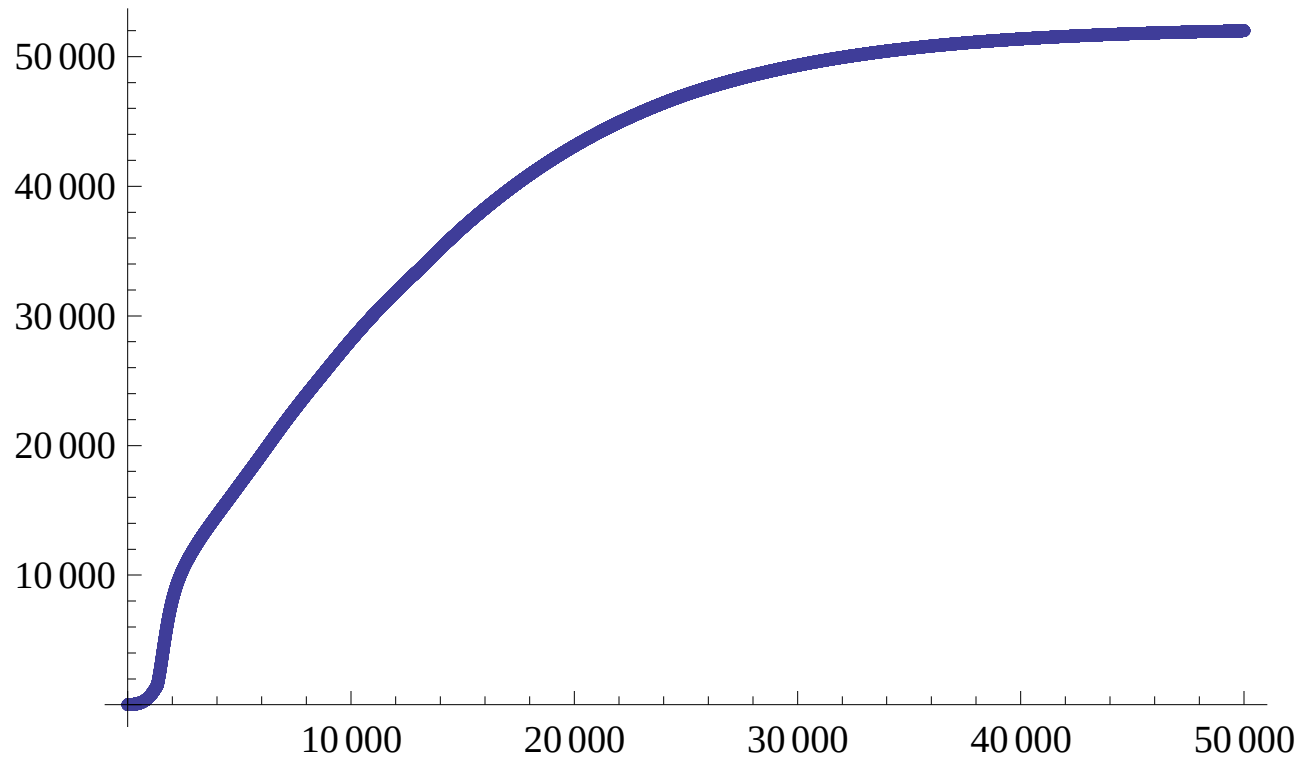

**We rearrange the x – axis to correspond to time in sec.**

```
Table[{10^(-5) i, %53[[i]]}, {i, 1, 50 001}];
```

```
ListPlot[%70, PlotRange → All]
```

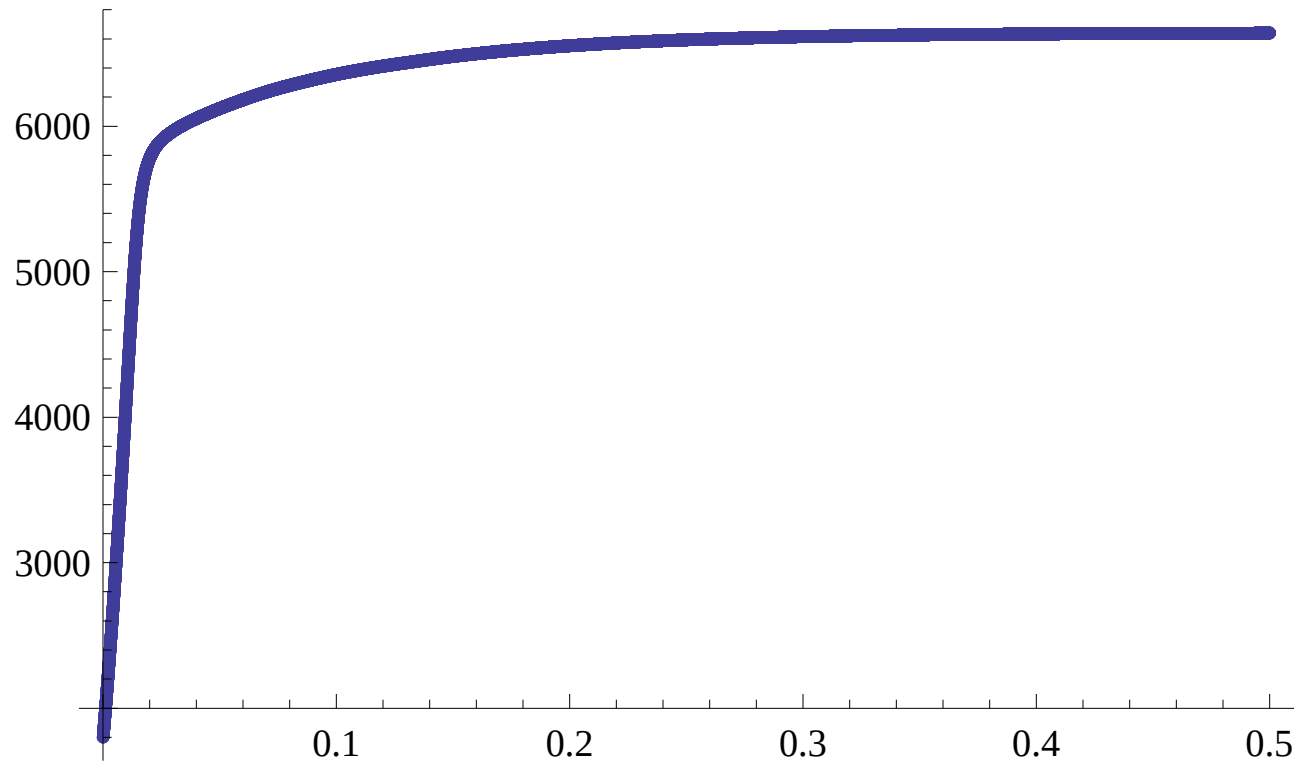

**Watt calculates the continuum radiation.**

```
Watt[U_] := Sum[(4 Pi / 3) If[i == 1, W[U[[1]][[2]]] U[[1]][[1]]^3,
    W[U[[i]][[2]]] (U[[i]][[1]]^3 - U[[i - 1]][[1]]^3)], {i, 1, 30}]
```

```
Watt[%55]
```

0.908783

# Supplementary Material - Foo Fighters.

**Here we do the same thing for a larger ball.**

```
w1 = Interpolation[{{0, 0}, {1000, 1.5}, {1500, 4.5}, {2000, 15}, {3000, 150},
   {5000, 1500}, {10 000, 7.47 × 10^7}, {12 500, 10^9}, {20 000, 7 × 10^9}, {30 000, 10^10}}]
```

InterpolatingFunction[{{0., 30 000.}}, <>]

```
w2[t_] := Exp[.00216329 t - 3.50388]
```

```
W[t_] := If[5000 < t < 10 000, w2[t], w1[t]]
```

```
d0 = .1
k = 1.38 × 10^-23
β = 3.5
η0 = 2.45 × 10^25
T0 = 1800
P0 = 101 325
I0 = 250 000
```

```
N0 = 4 (300 / 1800) Pi η0 d0^3 / 3
```

$1.71042 \times 10^{22}$

```
V0 = 4 Pi d0^3 / 3
```

0.00418879

```
N0 / V0
```

$4.08333 \times 10^{24}$

```
Pi I0 / (N0 k 3.5)
```

```
950 691.
```

```
HH[U_] :=
 Table[{i, Which[i == 1, 2 (U[[2]][[2]] - U[[1]][[2]]) / U[[1]][[1]]^2, i == 30, 0, i == 31, 0, True,
    2 ((U[[i + 1]][[2]] - U[[i]][[2]]) / (U[[i + 1]][[1]] - U[[i]][[1]]) - (U[[i]][[2]] -
           U[[i - 1]][[2]]) / (U[[i]][[1]] - U[[i - 1]][[1]])) / (U[[i + 1]][[1]] - U[[i - 1]][[1]]) +
     2 (U[[i]][[2]] - U[[i - 1]][[2]]) / ((U[[i]][[1]] - U[[i - 1]][[1]]) U[[i]][[1]])]}, {i, 1, 30}]
```

```
G[U_, M0_, i_] := (U[[i]][[2]] / M0[[i]][[2]])
```

```
D0 = Interpolation[{{300, .000022}, {1000, .00015}, {2000, .00052},
   {3000, .0024}, {4000, .0043}, {5000, .009}, {6000, .027}, {7000, .053},
   {10 000, .024}, {15 000, .065}, {20 000, .093}, {25 000, .195}, {30 000, .3}}]
```

```
InterpolatingFunction[{{300., 30 000.}}, <>]
```

**Here we use a larger .001 sec time step.**

```
Diff[U_] := Table[If[i ≤ 30, {U[[i]][[1]], U[[i]][[2]] + .001 D0[U[[i]][[2]]] HH[U][[i]][[2]]},
   U[[i]] + .001 {1, 1}], {i, 1, 31}]
```

**Here we leave the source on always.**

```
Source[U_] :=
 Table[Which[i ≤ 10, .001 {0, If[U[[31]][[2]] ≤ 1, 1, 0] 950 691} (U[[10]][[1]])^2 + U[[i]] -
    {0, 1} .001 U[[i]][[2]] W[U[[i]][[2]]] / (P0 β), True, {0, 0} + U[[i]]], {i, 1, 31}]
```

```
K = Table[Which[i ≤ 10, {.01 i, 1800}, 10 < i ≤ 30, {.01 i, 300}, True, {0, 0}], {i, 1, 31}]
```

```
{{0.01, 1800}, {0.02, 1800}, {0.03, 1800}, {0.04, 1800}, {0.05, 1800}, {0.06, 1800},
 {0.07, 1800}, {0.08, 1800}, {0.09, 1800}, {0.1, 1800}, {0.11, 300}, {0.12, 300},
 {0.13, 300}, {0.14, 300}, {0.15, 300}, {0.16, 300}, {0.17, 300}, {0.18, 300},
 {0.19, 300}, {0.2, 300}, {0.21, 300}, {0.22, 300}, {0.23, 300}, {0.24, 300},
 {0.25, 300}, {0.26, 300}, {0.27, 300}, {0.28, 300}, {0.29, 300}, {0.3, 300}, {0, 0}}
```

```
HD[U_] := Diff[Source[U]]
```

```
FF[U_, M0_] := Table[{(∑_{k=1}^{j} (G[U, M0, k] (U[[k]][[1]]^3 - If[k > 1, 1, 0] U[[k - 1]][[1]]^3)))^(1 / 3),
    U[[j]][[2]]}, {j, 1, 31}] // N
```

```
Evolve[U_] := FF[HD[U], U]
```

```
HD[K]
```

```
{{0.01, 1807.61}, {0.02, 1807.61}, {0.03, 1807.61}, {0.04, 1807.61}, {0.05, 1807.61},
 {0.06, 1807.61}, {0.07, 1807.61}, {0.08, 1807.61}, {0.09, 1807.61}, {0.1, 1801.94},
 {0.11, 300.271}, {0.12, 300}, {0.13, 300}, {0.14, 300}, {0.15, 300}, {0.16, 300}, {0.17, 300},
 {0.18, 300}, {0.19, 300}, {0.2, 300}, {0.21, 300}, {0.22, 300}, {0.23, 300}, {0.24, 300},
 {0.25, 300}, {0.26, 300}, {0.27, 300}, {0.28, 300}, {0.29, 300}, {0.3, 300}, {0.001, 0.001}}
```

```
NestList[Evolve, K, 501];
```

```
CT[U_] := U[[1]][[2]]
```

```
Table[CT[%31[[i]]], {i, 1, 501}];
```

```
ListPlot[%]
```

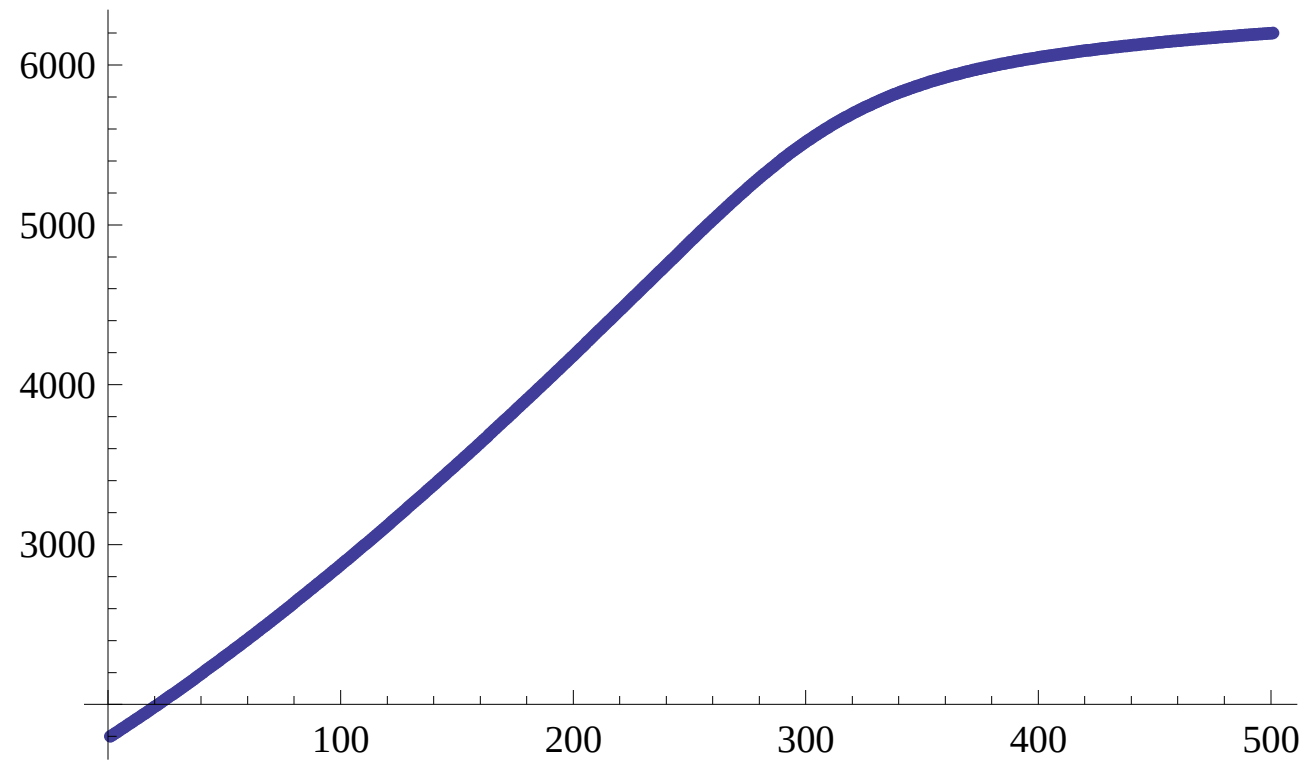

**Here is the temperature profile at .5 sec.**

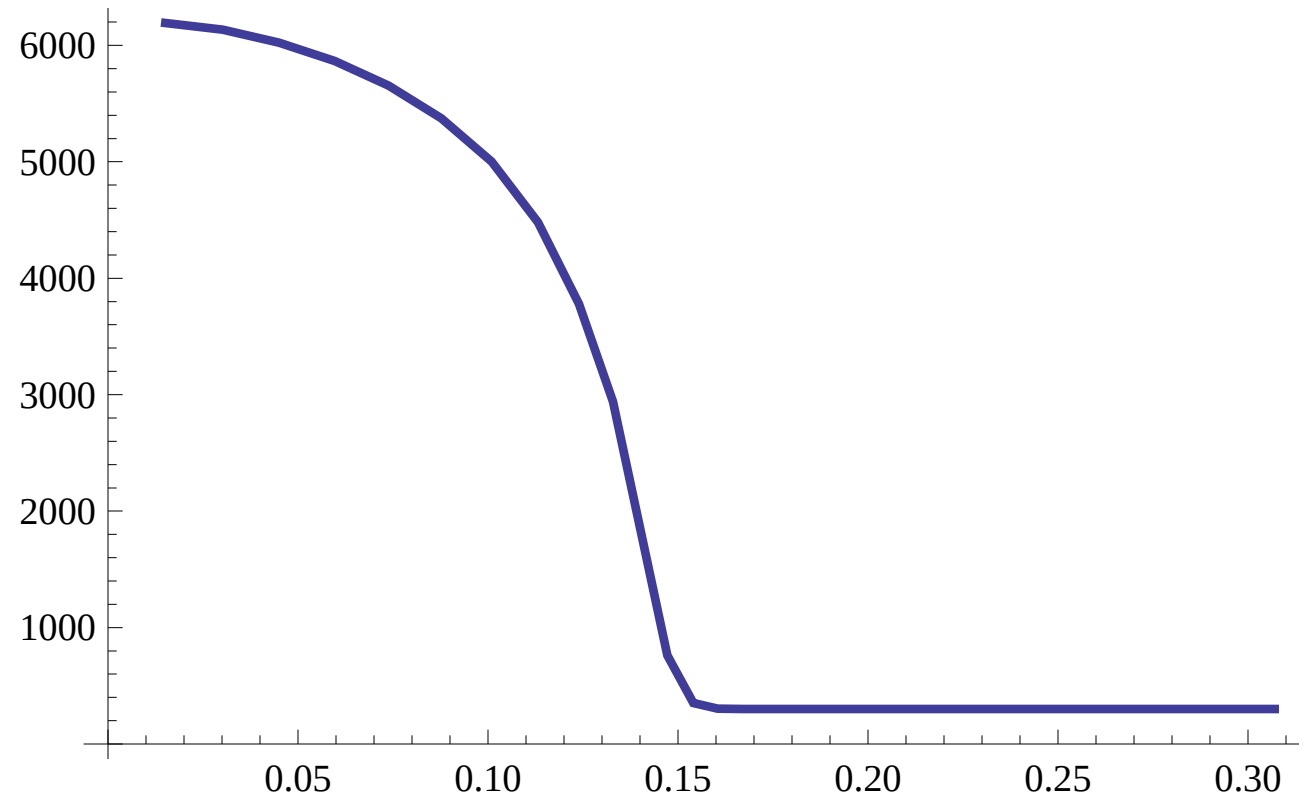

**Here we turn the source off after .35 sec.**

```
Source[U_] :=
 Table[Which[i ≤ 10, .001 {0, If[U[[31]][[2]] ≤ .35, 1, 0] 950 691} (U[[10]][[1]])^2 + U[[i]] -
   {0, 1} .001 U[[i]][[2]] W[U[[i]][[2]]] / (P0 β), True, {0, 0} + U[[i]]], {i, 1, 31}]
```

```
NestList[Evolve, K, 501];
```

```
Table[CT[%75[[i]]], {i, 1, 501}];
```

```
ListPlot[%]
```

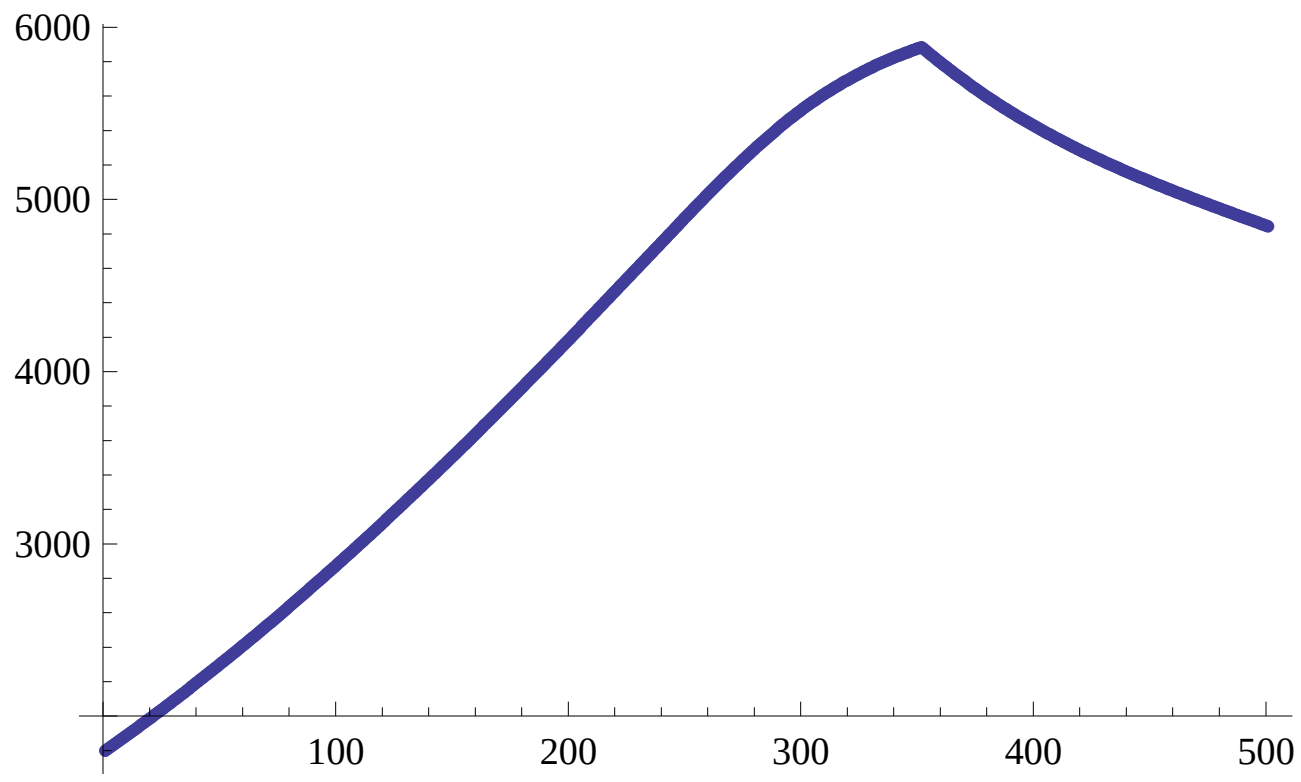

```
Watt[U_] := Sum[(4 Pi / 3) If[i == 1, W[U[[1]][[2]]] U[[1]][[1]]^3,
    W[U[[i]][[2]]] (U[[i]][[1]]^3 - U[[i - 1]][[1]]^3)], {i, 1, 30}]

Table[Watt[%32[[i]]], {i, 1, 501, 10}];
```

**Below we plot the various balls' luminosity as a function of their distance from the aircraft.**

```
ListPlot[%]
```

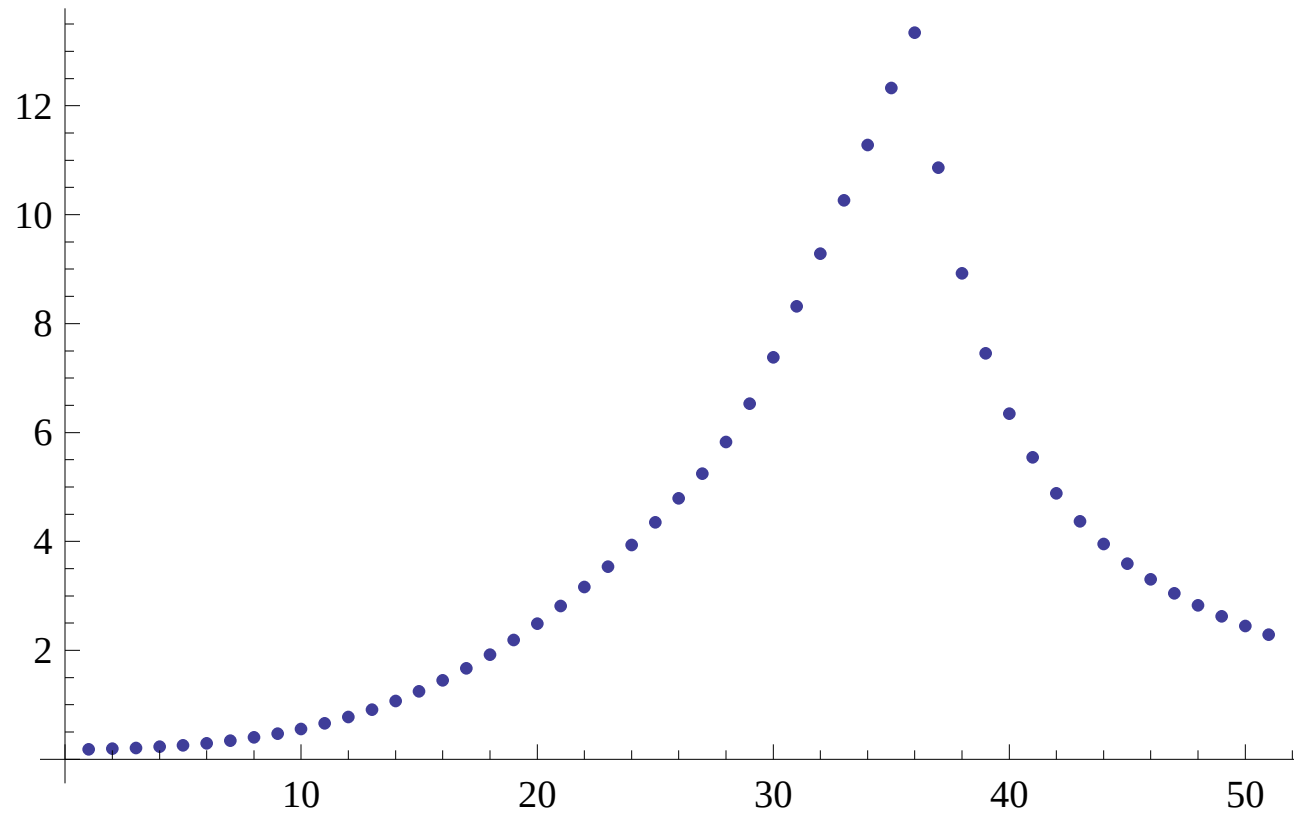